\documentstyle[preprint,aps,prl,epsf]{revtex}
\begin{document}
\def\bea{\begin{eqnarray}}
\def\eea{\end{eqnarray}}
\def\a{\alpha}
\def\d{\delta}
\def\p{\partial}
\def\nn{\nonumber}
\def\r{\rho}
\def\rv{\bar{r}}
\def\la{\langle}
\def\ra{\rangle}
\def\e{\epsilon}
\def\o{\omega}
\def\n{\eta}
\def\g{\gamma}
\def\break#1{\pagebreak \vspace*{#1}}
\def\f{\frac}
\twocolumn[\hsize\textwidth\columnwidth\hsize\csname
@twocolumnfalse\endcsname
\draft
\title{Dynamics of Vibrated Grains}
\author{Abhishek Dhar$^{1}$ \cite{ABHI} and Supurna Sinha$^{2}$\cite{SUP}}
\address{Raman Research Institute,
Bangalore 560080,India\\}
\maketitle
\widetext
\begin{abstract}
We study number density distribution and the behavior of time correlation 
functions in the density of grains for a quasi-two dimensional system of 
vibrated grains. 
We study the system at various 
packing fractions, from low to high. At low densities we recover usual 
gas like behavior, reflected in a Poissonian statistics for the number 
density distribution. At higher densities we notice 
effects like formation of cages of the kind 
that are seen in glass transition. 
We study these effects 
with a perspective of understanding the similarities and differences
between an atomic fluid and a ``scaled up fluid'' like a vibrated granular
system. 
\end{abstract}

\pacs{}]
\narrowtext
In recent years, granular systems have emerged as an active area of 
research. Interest in this field has grown as a result of observations
coming from interesting and relatively low-tech experiments being done
around the world \cite{prevost}. 
A vibrated granular system consisting of a large number of macroscopic
grains in rapid motion, provides us with a large-scale 
statistical mechanical system which is of interest both for its
relevance to fundamental theoretical issues such as fluctuation-dissipation
connection, notion of temperature and so on and its importance in 
industrial applications.

Here we have experimentally studied a quasi two dimensional granular
system consisting of a layer of a large number of spherical particles
(mustard seeds) covering a vertically driven horizontal ground glass plate.
The driving was provided by a speaker attached to a signal generator.
The chosen frequency setting was $300 Hz$. We carried out the experiment
at various densities ; typical values being $400$, $600$ and 
$800$ particles in an area  of $9.5\times12$ $sq$ $mm$. 
In order to analyze the number density 
distributions and the time dependent number density correlation functions
we had the following set up. We captured each visual frame of the vibrated 
grains using a CCD camera attached to a video system. We could see the movie
on a monitor\cite{movie}. Each frame was tagged using a timer which kept 
track
of the time of recording. Subsequently we converted the film frames recorded
in the videotape into $bmp$ files in the computer and analyzed and plotted
our data. 

Our low packing fraction data
fitted well to a Poissonian statistical distribution for the number 
distribution for the grains (See Fig. 1). 
This indicates that at low densities a vibrated
granular system has a dilute gas like behavior. We expect a better
agreement with Poissonian statistics in the dilute limit
for a larger number of data points. 
As we probed higher
densities we noticed the formation of cages. We intend to study 
this caging with a view to undestanding 
the formation of glassy 
states in such systems \cite{glass}. 
One important difference between probing a liquid consisting 
of smaller sized particles, say, a colloidal glass forming liquid
and probing glass-like states in such vibrated granular system is that we 
can study these effects in great detail by directly looking at the system
without the use of a microscope. In colloids only recently due
to advances in technology researchers can probe cage formations by 
looking through a confocal microscope \cite{weeks}.
In order to probe such 
effects in detail we need to study the behavior of velocity autocorrelation
functions which we expect to reflect back-scattering effects due to cage 
formation. Such back scattering effects would lead to negative velocity
autocorrelations. It would be interesting to look at tagged particle 
diffusion and 
study the slowing down of diffusional relaxation with the increase in the
packing fraction in such a system.
Such a study is expected to shed light on 
the similarities and differences
between an atomic fluid and a ``scaled up fluid'' like a vibrated granular
system. 
\vspace{-0.5cm}
\vbox{
\epsfxsize=9.0cm
\epsfysize=10.0cm
\epsffile{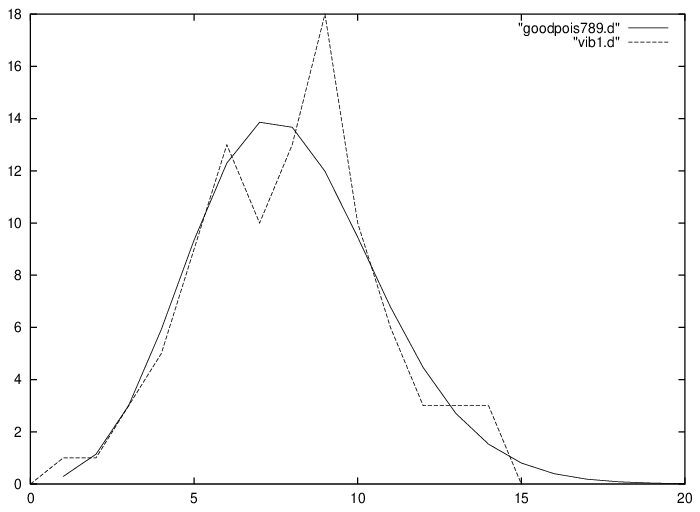}
\begin{figure}
\caption{The number distribution of vibrated grains (dashed line) for
$400$ grains contained in an area of $8\times 8$ $sq$ $cm$
compared against a Poissonian distribution (solid line) with the 
same mean value as the experimentally determined one. This experimental
run involved $100$ data points.} 
\label{dist}
\end{figure}}
{\it Acknowledgements:}
We thank R. Pratibha for providing us with a CCD camera and a video
equipment and P. Vishwanath for helping us with digital conversion of 
data. The work based on this technical report was done at the Theorists' 
Laboratory at RRI with help from the RRI workshop.

\end{document}